# A Synthetic Skyrmion Platform with Robust Tunability

Yong Li,[1,#] Qiyuan Feng,[2,3,#] Sihua Li,[4,#] Ke Huang,[4] Weiliang Gan,[4] Haibiao Zhou,[2,3] Xiangjun Jin,[1] Xiao Renshaw Wang,[4,5] Fusheng Ma,[1,*] Qingyou Lu,[2,3,*] and Wen Siang Lew[4,*]

[1]*Jiangsu Key Laboratory of Opto-Electronic Technology, Center for Quantum Transport and Thermal Energy Science, School of Physics and Technology, Nanjing Normal University, Nanjing 210046, China*
[2]*Anhui Province Key Laboratory of Condensed Matter Physics at Extreme Conditions, High Magnetic Field Laboratory, Chinese Academy of Sciences, Hefei, Anhui, China.*
[3]*Hefei National Laboratory for Physical Sciences at the Microscale, University of Science and Technology of China, Hefei, Anhui, China.*
[4]*School of Physical and Mathematical Sciences, Nanyang Technological University, 21 Nanyang Link, Singapore 637371*
[5]*School of Electrical and Electronic Engineering, Nanyang Technological University, 50 Nanyang Ave, Singapore 639798*

#These authors contribute equally to this work
*Corresponding Email: phymafs@njnu.edu.cn; qxl@ustc.edu.cn; wensiang@ntu.edu.sg;

## Abstract

Magnetic skyrmions are topologically non-trivial spin structure, and their existence in ferromagnetically coupled multilayers has been reported with disordered arrangement. In these multilayers, the heavy metal spacing layers provide an interfacial Dzyaloshinskii-Moriya interaction (DMI) for stabilizing skyrmions at the expense of interlayer exchanging coupling (IEC). To meet the functional requirement of ordered/designable arrangement, in this work, we proposed and experimentally demonstrated a scenario of skyrmion nucleation using nanostructured synthetic antiferromagnetic (SAF) multilayers. Instead of relying on DMI, the antiferromagnetic IEC in the SAF multilayers fulfills the role of nucleation and stabilization of skyrmions. The IEC induced skyrmions were identified directly imaged with MFM and confirmed by magnetometry and magnetoresistance measurements as well as micromagnetic simulation. Furthermore, the robustness of the proposed skyrmion nucleation scenario was examined against temperature (from 4.5 to 300 K), device size (from 400 to 1200 nm), and different lattice designs. Hence, our results provide a synthetic skyrmion platform meeting the functional needs in magnonic and spintronic applications.

# Introduction

Magnetic skyrmions are topologically protected quasi-particle with non-trivial spin structure, which have been observed experimentally in non-centrosymmetric bulk magnetic materials[1–8] such as B20 compound MnSi, FeGe and PdFe. Recently, stabilization of nanoscale skyrmions at room temperatures (RT) were also experimentally realized in ultrathin Ir/Fe,[9] Ta/CoFeB/MgO,[10] Ta/CoFeB/TaO,[10] Pt/Co,[11,12] Ir/Co/Pt,[12] Pt/Co/MgO[13] and Ir/Fe/Co/Pt[14] multilayer films. In these multilayers, an asymmetric Dzyaloshinskii-Moriya interaction (DMI)[15,16] appears at the ferromagnet/heavy-metal interface and stabilizes the random existence of skyrmion. Benefiting from the nanoscale size, topologically protected stability, and easy current-driven motion, magnetic skyrmions are highly promising as an information carrier in next-generation spintronic devices for data storage and logic operation. Alternatively, artificially periodic arrangements of skyrmions in 1D or 2D provides a novel type of 'metamaterial', *i.e.* skyrmion-based magnonic crystals (SBMCs),[17–20] in which the magnetization can be periodically modulated by ordered skyrmion lattices. In comparison with lithographically patterned magnonic crystals, a strong advantage of SBMCs is that the propagation of spin waves inside can be dynamically manipulated by simply reconfiguring the property of skyrmion.[17] Furthermore, SBMCs are suitable candidates for topological matter investigation, such as the realization of topological magnonic insulators.[18–20] However, the realization of SBMCs is restricted by the random nucleation of skyrmion in magnetic multilayers with interfacial DMI.[10–14] One route to realize ordered skyrmion arrangement, which allows for high/flexible tunability and exploits artificial skyrmions, have recently been proposed and realized at room temperature. [21–29] These artificial skyrmions are realized by imprinting the magnetic vortex of the top nanodots into the underlayer film with perpendicular magnetic anisotropy (PMA) without the requirement of DMI. Although the arrangement of skyrmion can be flexibly designed by nanopatterning the top nanodots, the skyrmion size cannot be adjusted as they are geometrically confined within the area under the top nanodots. A skyrmion platform, showing arrangement and size tunability simultaneously, is still missing.

Recently, the spin texture of skyrmions in multilayers has been demonstrated to be non-identical for all the individual magnetic layers, and hence they cannot be effectively described as a 2D spin texture as shown in Figure 1a.[30,31] Considering a ferromagnet/non-magnet/ferromagnet system, the two magnetic layers can effectively interact via either interlayer dipolar coupling (IDC) or IEC depending on the thickness of the non-magnetic spacing layer. As shown in Figure 1a, the IDC[10–14] and ferromagnetic IEC[28,32] stabilize hybrid skyrmions with the same topology; while the antiferromagnetic (AFM) IEC nucleates skyrmions of opposite topology charge. [33]

In this work, we propose and experimentally demonstrate the nucleation of artificial skyrmions in nanostructured synthetic antiferromagnetic (SAF) multilayer without requiring of DMI. The AFM IEC between the top nanodots and the film beneath gives rise to the stabilization of skyrmions. Once it is nucleated, the skyrmion size can be adjusted and is not limited by the geometrical confinement of the top nanodots. The synthetic skyrmion formation process is characterized by MFM and further confirmed with magnetometry and magnetoresistance measurements as well as micromagnetic simulations. Furthermore, the robustness of the skyrmion nucleation mechanism against temperature (from 4.5 to 300 K) and device size (from 400 to 1200 nm) are also examined. The demonstrated synthetic skyrmion system provides a readily designable platform for prospective materials based on ordered skyrmion arrangement.

## Results

Figure 1b shows the cross-sections of skyrmion spin textures along the radial directions in nanostructured SAF multilayers, in which the top ferromagnetic (FM) layer is patterned into 2D arrays of circular dots sand the bottom FM layer is a continuous film. Considering the FM IEC, an skyrmion-like spin texture can be formed across both the top dot and bottom film as reported previously.[28,34–36] Our SAF multilayer is non-compensated and the IEC is antiferromagnetic type. As shown in Figure 1b, there are three possible scenarios where skyrmions can exist: 1). skyrmion exists in both the top dot and the bottom film but an opposite topology charge; 2). skyrmion exists only in the top dot; 3). skyrmion exists only in the bottom film beneath the top dot. These three types of skyrmions can be distinguished from their corresponding MFM images as simulated in Figure 1b.

The used SAF multilayers are magnetron-sputtered in the non-compensated form of Ta(4)/Pt(4)/[Pt(0.6)/Co(0.6)]$_2$/Ru/[Co(0.6)/Pt(0.6)]$_4$/Ta(4) (the number in parentheses are nominal thickness in nm, will be indicated as [Pt/Co]$_2$/Ru/[Co/Pt]$_4$). By tuning the thickness of Ru spacer, the top [Co(0.6)/Pt(0.6)]$_4$ layers and the bottom [Pt(0.6)/Co(0.6)]$_2$ layers are antiferromagnetically coupled through the IEC. For reference, three Ta(4)/Pt(4)/[Pt(0.6)/Co(0.6)]$_n$/Ta(4) multilayers ($n$ = 2, 4, and 6) and one compensated Ta(4)/Pt(4)/[Pt(0.6)/Co(0.6)]$_4$/Ru/ [Co(0.6)/Pt(0.6)]$_4$/Ta(4) SAF multilayers were also prepared. The magnetization reversal properties of the SAF multilayer and reference multilayers are characterized showing multi- and single-step switching, see details in Supplementary Information (SI). The magnetization hysteresis (*M-H*) loop of the [Pt/Co]$_2$/Ru/[Co/Pt]$_4$ multilayer exhibits a typical antiferromagnetic characteristics indicating the presence of AFM IEC. To experimentally demonstrate the proposed skyrmion nucleation scenario with AFM IEC, the top [Co/Pt]$_4$ layer was patterned into circular dots of different diameters ranging from 400 to 1200 nm. In such a half-etched structure, the bottom continuous [Pt/Co]$_2$ film can be divided into two regions: one is the non-dot-covered region behaving as ferromagnetic films; the other one is the dot-covered region behaving as SAF together with the top dot.

The room temperature magnetic force microscopy (MFM) measurement was carried out with an out-of-plane magnetic field varying from +12 to -12 kOe. Figure 2a shows the historical changing of the magnetic morphology of the nanostructured SAF multilayers at selected fields. The diameters of the top [Co/Pt]$_4$ dots changes from 400 nm (first row) to 1200 nm (fifth row) with an interval of 200 nm. Their physical properties can be found from the SEM images in SI as shown in Figure S1. It is observed that the magnetization changed from positively saturated state to non-saturated state to negatively saturated state. Interestingly, in the non-saturated state, the dot-covered region exhibits different magnetic morphology. Generally, the dot-covered region exhibits a brighter color and shrinks inward until eventually annihilated with field decreasing. For instance, at the field of -2.6 and -2.7 kOe, the MFM images of the dot with diameter of 600 nm exhibit a white spot at the second row of Figure 2a. This observation is different form the reported artificial skyrmion, whose sizes cannot be adjusted and limited by the geometrical size of the top dot.[21–29]

To understand what type of skyrmion corresponding to the experimental MFM images, we performed micromagnetic simulations to calculate the MH loops, spin textures, and MFM images of nanostructured SAF multilayer with a top dot of diameter 400 nm as shown in Figure 2b. From the simulated *M-H* loop, five magnetization stages are observed with field decreasing from +6 to -6 kOe. By analyzing the spin textures at each magnetization stage, see insets in Figure 2b, the whole magnetization reversal process can be understood as: stage 1-positively saturated; stage 2-top dot switched by AFM IEC; stage 3-non-dot-covered region of the bottom layer switched by the reversed magnetic field, while dot-covered region not switched as protected by AFM IEC; stage 4-dot-covered region shrinks as the increasing negative field are able to gradually offset the AFM IEC protection;

and stage 5-negatively saturated. Therefore, a skyrmion spin texture is realized in the bottom continuous film at both stage 3 and stage 4. The skyrmion at stage 3 has fixed size limited by the geometry of the top dot; while the size of skyrmion at stage 4 can be adjusted by changing field. By comparing the measured and simulated MFM images, the states marked by five dashed rectangles at the second row of Figure 2a are one-to-one matched with the simulated stages as indicated in Figure 2b, we can preliminarily confirmed that the experimentally observed white spots in the MFM images indicate the successful appearance of AFM IEC induced skyrmion. To clarify this point further, we did a line-cut along the dot radial direction (red dash line) of the measured and simulated MFM images at state 4 as shown in Figures 2c and d, respectively. The experimental MFM image was marked by dashed circle at the first row of Figure 2a. While for the simulated MFM image, a tip height of 50 nm is chosen with the best agreement with experimental MFM images. The detailed comparison of simulated MFM images under different tip heights can be found in SI as shown in Figure S2. There is a good consistency between the measured and simulated MFM images in Figures 2c,d. Therefore, we can conclude that the observed AFM IEC skyrmion is the case that skyrmion exists only in the bottom film beneath the top dot, whose cross-section view and 3D view are shown in Figures 1b and 2e, respectively. To be noted that we also did the MFM measurement on a fully etched SAF multilayer sample, i.e. both the top and bottom layers are patterned into a circular dot. The AFM IEC skyrmion corresponded MFM images were not observed as shown in Figure S3.

The nucleation scenario of AFM IEC skyrmion has been demonstrated by direct MFM imaging. We also employ two indirect methods to confirm the imaging observations: *M-H* loop measurement by a polar magneto-optical Kerr effect (MOKE) technique and anomalous Hall effect (AHE) measurements by 4-point magneto-transport technique. The *M-H* and anomalous Hall resistance ($R_{AHE}$–H) loops of nanostructured SAF multilayers at RT with a top dot diameter of 400 nm are shown in Figures 3a and b, respectively. For the optical measurement of *M-H* loops, a 2D arrangement of 400 nm dots in a square lattice with an area of 800 nm was prepared as the inserted SEM images in Figure 3a. The schematic of one unit cell is also displayed with a period of 800 nm. Consistent with the simulated *M-H* loop in Figure 2b, the optically measured *M-H* loop also exhibits five magnetization stages, in which the AFM IEC skyrmion stage is highlighted. Besides measuring major *M-H* loops, we also carried out the minor *M-H* loop measurement, in which the field decreases from a positively saturated value $+H_{sat}$ to a selected reversal field $H_R$ and then back to $+H_{sat}$. The selection of different $H_R$ (50 Oe, -1500 Oe, and -2500 Oe) is to choose different spin texture states (stage 2, stage 3, and stage 4) as seen from the insets in Figure 3c. It is clearly found that the minor loops exhibit an irreversible magnetization process. For the condition of $H_R$ = -2500 Oe and -1500 Oe, the bottom layer is firstly saturated by external field through skyrmion expanding and reach the spin texture state at $H_R$ = 50 Oe. For $H_R$ = 50 Oe, the AFM IEC protects the top dot layer from saturating by the external field and resulting in a much higher saturation field +$H_{sat}$ than that of the major loop. This irreversible magnetization process also indicates that the spin texture at stage 2 is a purely AFM state (top dot and bottom film are saturated in opposite directions) rather than an AFM IEC skyrmion state. If the spin texture at stage 2 is in the form of AFM IEC skyrmion, the minor reversal loop will consistent with the major loop as schematically explained in Figure S4. Figure 3b/d is the measured major/minor $R_{AHE}$–H loop of the nanostructured SAF multilayer with a top dot of diameter 400nm at room temperature. The bottom [Pt/Co]$_2$ layer was fabricated into cross Hall bar with a width of 3 μm, and the [Co/Pt]$_4$ layer was patterned into a circular dot of diameter 400 nm located at the center of the Hall bar. The $R_{AHE}$ behaves in a similar manner to the optically measured magnetization. Hence, both the *M-H* loop and $R_{AHE}$–H loop measurement can act as an indirect method to characterize the presence of AFM IEC skyrmion.

So far, we have mainly focused on the demonstration of creating the proposed AFM IEC skyrmion at room temperature and nanostructured the SAF multilayer with a top dot of diameter of 400 nm creation. To exam the generality and robustness of the scenario, we expand the investigation to smaller dot-sizes and lower temperatures. Figure 4a shows the optically measured *M-H* loops of nanostructured SAF multilayers with top dots of diameters 200 nm, 300 nm, 400 nm, 600 nm and arranged in a square lattice. It is found that the saturation field and skyrmion existing field range increase slightly with dot size increasing, while the switching field of the bottom layer decreases with dot size increasing. To investigate the effect from the dots arrangement, we also fabricated a honeycomb arranged dots of 400 nm diameter. There is no difference between the *M-H* loops of the square and honeycomb lattice arranged dots. Figure 4b shows the $R_{AHE}$-H loops for three top dot sizes (200 nm, 300 nm and 400 nm). A similar phenomenon to *M-H* loops is observed that the top dot size slightly affect the skyrmion existing field ranges. The skyrmion annihilation field or the saturation field decreases with dot size decreasing. Therefore, comprehensively considering the MFM measurement in Figure 2a, the *M-H* and $R_{AHE}$–H loops in Figure 4, the nucleation of skyrmion is robust against the dot size with a slight variation of the skyrmion existing field range.

From the application point of view, the temperature-dependent stability is also an important factor, particularly for spintronic space devices. To exam the stability of skyrmion nucleation, we did the $R_{AHE}$–H loop and MFM measurements at low temperatures ranging from 4.5 to 300 K. Figure 5a shows the fraction of the $R_{AHE}$–H loops corresponding to the skyrmion existing field range for the top dot of 400 nm diameter, it is found that the skyrmion field ranges (from the starting of stage 3 to the annihilation of stage 4) increase to higher fields with temperature decreasing. The temperature dependence of the skyrmion existing fields is more clearly described in the field-temperature diagram as shown in Figure 5b. This phase diagram is extracted from the low-temperature MFM images at 4.5 K, 100 K and 200 K for the top dot of 400 nm diameter as shown in Figure 5c,d,e, respectively. For easily understanding, we divide the diagram into three phases: reversal state (the below skyrmion size is equal to the physical size of the top dot), skyrmion state (the below skyrmion size is smaller than the physical size of the top dot) and saturation state. It is can found that the skyrmion state moves to higher field (also see dashed circles in Figure 5c-e) and becomes broader at a lower temperature. Together with the observation from the temperature-dependent $R_{AHE}$–H loops, this can be understood from the enhancement of the AFM IEC strength of the SAF multilayers at a lower temperature as shown in Figure S6.

## Conclusion

We experimentally demonstrated a synthetic skyrmion platform induced by the antiferromagnetic interlayer exchange coupling of the synthetic antiferromagnetic multilayers. The skyrmion nucleation scenario is robust to dimension, temperature, and field, and can be easily measured with direct magnetic imaging and indirect detection by magneto-optical and electron-transport technology. Our demonstrated IEC synthetic skyrmions greatly enrich the flexibility of skyrmionic devices and could also provide a platform for designing functional magnonic and spintronic structures based on ordered skyrmion arrangement.

# Methods:

**Sample preparation:** The [Pt(0.6)/Co(0.6)]$_2$/Ru(0.9)/[Co(0.6)/Pt(0.6)]$_4$ thin film were deposited on the thermally oxidized silicon wafer use DC magnetron sputtering technique at RT. Argon gas (~2.3 x 10$^{-3}$ Torr, 1 Torr = 1.33322 x 10$^2$ Pa) was used during the sputtering process with a background pressure of 2 × 10$^{-8}$ Torr, and the deposition rates for Pt, Co, and Ru were 0.14, 0.21, and 0.10 Å s$^{-1}$, respectively. In the thin film stack, the bottom stack [Pt(0.6)/Co(0.6)]$_2$ and the top stack [Co(0.6)/Pt(0.6)]$_4$ are antiferromagnetically coupled through a Ru spacer layer. For contrast, we patterned the multilayer into different nanodots structures by electron beam lithography (EBL) and Ar$^+$ ion milling methods. The Hall bar is 40 μm long and 3 μm wide.

**Kerr imaging and *M-H* loops:** Polar MOKE measurement was used to characterize the magnetic properties. The MagVision Kerr microscopy system operating in differential imaging mode was used to produce the magnetization hysteresis (both major and minor) loops. An ultrabright 525 nm light source allowed for data capture rates up to 60 Hz.

**Magnetic Force Microscopy experiments:** The MFM experiments were performed using a home-built variable temperature MFM, equipped with a 20 T superconducting magnet.[37,38] We incorporated a commercial piezoresistive cantilever (PRC400; Hitachi High-Tech Science Corporation) as the force sensor. The resonant frequency of the cantilever is about 42 kHz. The MFM tip has a magnetic coating consisting of 5 nm Cr, 50 nm Fe, and then 5 nm Au films. This magnetic coating was magnetized perpendicular to the cantilever. The magnetic coercivity and saturation fields are ~250 Oe and ~2000 Oe, respectively. A built-in phase-locked loop (R9 controller; RHK Technology) was utilized for MFM scanning control and signal processing. MFM images were collected in a constant height mode. First, a topographic image was obtained using contact mode, from which the sample surface tilting along the fast and slow scan axes could be compensated. Then the tip was lifted by ~100 nm to the surface and MFM images were obtained in frequency-modulation mode.

**Anomalous Hall resistance measurement:** The $R_{AHE}$ were carried out on a Cascade Microtech probe station. The Hall resistance was determined by measuring the voltage between the Hall bar using a multimeter (Keithley 2082A) while applying a DC current of 100 μA using a sourcemeter (Keithley 6221).

**Micromagnetic Simulations:** The micromagnetic simulation was carried out using a commercial Landau Lifschitz Gilbert (LLG) Micromagnetic Simulator.[39] We focus on simulating multilayer nanodots system which the top FM layer is dot structure with 400 nm diameter and the bottom FM layer is the square structure with 2 μm × 2 μm × 1 nm size. The cell size of all layers is 5 nm × 5 nm × 1nm. The material parameters used in this simulation were saturation magnetization: $M_{FM}$ = 1020 emμ/cm$^3$; exchange constant $A$ = 1.05 μerg/cm; the PMA constant $K_u$ = 7.6 × 10$^6$ erg/cm$^3$; the interlayer exchange coupling between two FM layer across nonmagnetic layer, $J_{iec}$ = -0.4 erg/cm$^2$, negative value indicate antiferromagnetic interlayer exchange coupling. The MFM images are also simulated with different tip height to compare with experiments.

**Acknowledgement**

F.M. acknowledges supports from the National Natural Science Foundation of China (Grant No. 11704191), the Natural Science Foundation of Jiangsu Province of China (Grant No. BK20171026), the Jiangsu Specially-Appointed Professor, and the Six-Talent Peaks Project in Jiangsu Province, China (Grant No. XYDXX-038). X.R.W. acknowledges supports from the Nanyang Assistant Professorship grant from Nanyang Technological University and Academic Research Fund Tier 1 (RG108/17 and RG177/18) and Tier 3 (MOE2018-T3-1-002) from Singapore Ministry of Education. Q.L. and Q.F. were supported by the National Key R&D Program of China (grants no. 2017YFA0402903 and no. 2016YFA0401003) and the National Natural Science Foundation of China (grant no. 51627901). W.S.L. acknowledge supports from a NRF-CRP Grant (CRP9-2011- 01), a RIE2020 ASTAR AME IAF-ICP Grant (No.I1801E0030) and an ASTAR AME Programmatic Grant (No. A1687b0033) are also acknowledged. W.S.L is a member of the SG-SPIN Consortium.

## References:

[1] A. Neubauer, C. Pfleiderer, B. Binz, A. Rosch, R. Ritz, P. G. Niklowitz, P. Böni, *Phys. Rev. Lett.* **2009**, *102*, 1.

[2] M. C. Langner, S. Roy, S. K. Mishra, J. C. T. Lee, X. W. Shi, M. A. Hossain, Y. Chuang, S. Seki, Y. Tokura, S. D. Kevan, R. W. Schoenlein, *Phys. Rev. Lett.* **2014**, *112*, 1.

[3] S. X. Huang, C. L. Chien, *Phys. Rev. Lett.* **2012**, *108*, 1.

[4] S. Mühlbauer, B. Binz, F. Jonietz, C. Pfleiderer, A. Rosch, A. Neubauer, R. Georgii, P. Böni, *Science .* **2009**, *323*, 915.

[5] X. Z. Yu, Y. Onose, N. Kanazawa, J. H. Park, J. H. Han, Y. Matsui, N. Nagaosa, Y. Tokura, *Nature* **2010**, *465*, 901.

[6] Y. T. s.seki, X.Z.Yu, S.Lshiwata, *Science* **2012**, *336*, 198.

[7] X. Z. Yu, N. Kanazawa, Y. Onose, K. Kimoto, W. Z. Zhang, S. Ishiwata, Y. Matsui, Y. Tokura, *Nat. Mater.* **2011**, *10*, 106.

[8] N. Romming, C. Hanneken, M. Menzel, J. E. Bickel, B. Wolter, K. Von Bergmann, A. Kubetzka, R. Wiesendanger, *Science.* **2013**, *341*, 636.

[9] S. Heinze, K. Von Bergmann, M. Menzel, J. Brede, A. Kubetzka, R. Wiesendanger, G. Bihlmayer, S. Blügel, *Nat. Phys.* **2011**, *7*, 713.

[10] W. Jiang, P. Upadhyaya, W. Zhang, G. Yu, M. B. Jungfleisch, F. Y. Fradin, J. E. Pearson, Y. Tserkovnyak, K. L. Wang, O. Heinonen, S. G. E. te Velthuis, A. Hoffmann, *Science.* **2015**, *11*, 1.

[11] S. Woo, K. Litzius, B. Krüger, M. Im, L. Caretta, K. Richter, M. Mann, A. Krone, R. M. Reeve, M. Weigand, P. Agrawal, I. Lemesh, M. Mawass, P. Fischer, M. Kläui, G. S. D. Beach, *Nat. Mater.* **2016**, *15*, 501.

[12] C. Moreau-Luchaire, C. Moutafis, N. Reyren, J. Sampaio, C. A. F. Vaz, N. Van Horne, K. Bouzehouane, K. Garcia, C. Deranlot, P. Warnicke, P. Wohlhüter, J. M. George, M. Weigand, J. Raabe, V. Cros, A. Fert, *Nat. Nanotechnol.* **2016**, *11*, 444.

[13] O. Boulle, J. Vogel, H. Yang, S. Pizzini, D. De Souza Chaves, A. Locatelli, T. O. Menteş, A. Sala, L. D. Buda-Prejbeanu, O. Klein, M. Belmeguenai, Y. Roussigné, A. Stashkevich, S. Mourad Chérif, L. Aballe, M. Foerster, M. Chshiev, S. Auffret, I. M. Miron, G. Gaudin, *Nat. Nanotechnol.* **2016**, *11*, 449.

[14] A. Soumyanarayanan, M. Raju, A. L. G. Oyarce, A. K. C. Tan, M.-Y. Im, A. P. Petrovic, P. Ho, K. H. Khoo, M. Tran, C. K. Gan, F. Ernult, C. Panagopoulos, *Nat. Mater.* **2017**, *16*, 898.

[15] I. Dzyaloshinsky, *J. Phys. C Solid State Phys.* **1958**, *4*, 241.

[16] Toru Moriya, *Phys. Rev. Lett.* **1960**, *4*, 4.

[17] F. Ma, Y. Zhou, H. B. Braun, W. S. Lew, *Nano Lett.* **2015**, *15*, 4029.

[18] A. Roldán-Molina, A. S. Nunez, J. Fernández-Rossier, *New J. Phys.* **2016**, *18*, 045015.

[19] S. A. Díaz, J. Klinovaja, D. Loss, *Phys. Rev. Lett.* **2019**, *122*, 187203.

[20] S. K. Kim, Y. Tserkovnyak, *Phys. Rev. Lett.* **2017**, *119*, 077204.

[21] L. Sun, R. X. Cao, B. F. Miao, Z. Feng, B. You, D. Wu, W. Zhang, A. Hu, H. F. Ding, *Phys. Rev. Lett.* **2013**, *167201*, 1.

[22] S. Pu, M. Dai, G. Sun, *Opt. Commun.* **2010**, *283*, 4012.

[23] B. F. Miao, L. Sun, Y. W. Wu, X. D. Tao, X. Xiong, Y. Wen, R. X. Cao, P. Wang, D. Wu, Q. F. Zhan, B. You, J. Du, R. W. Li, H. F. Ding, *Phys. Rev. B* **2014**, *90*, 174411.

[24] J. Li, A. Tan, K. W. Moon, A. Doran, M. A. Marcus, A. T. Young, E. Arenholz, S. Ma, R. F. Yang, C. Hwang, Z. Q. Qiu, *Nat. Commun.* **2014**, *4704*, 1.

[25] D. A. Gilbert, B. B. Maranville, A. L. Balk, B. J. Kirby, P. Fischer, D. T. Pierce, J. Unguris, J. A. Borchers, K. Liu, *Nat. Commun.* **2015**, *6*, 8462.
[26] K. Y. Guslienko, *IEEE Magn. Lett.* **2015**, *6*, 1.
[27] H. Z. Wu, B. F. Miao, L. Sun, D. Wu, H. F. Ding, *Phys. Rev. B* **2017**, *95*, 174416.
[28] G. Chen, A. Mascaraque, A. T. N. Diaye, A. K. Schmid, *Appl. Phys. Lett.* **2015**, *242404*, 1.
[29] Y. Y. Dai, H. Wang, P. Tao, T. Yang, W. J. Ren, Z. D. Zhang, *Phys. Rev. B* **2013**, *88*, 054403.
[30] W. Li, I. Bykova, S. Zhang, G. Yu, R. Tomasello, M. Carpentieri, Y. Liu, Y. Guang, J. Gräfe, M. Weigand, D. M. Burn, G. van der Laan, T. Hesjedal, Z. Yan, J. Feng, C. Wan, J. Wei, X. Wang, X. Zhang, H. Xu, C. Guo, H. Wei, G. Finocchio, X. Han, G. Schütz, *Adv. Mater.* **2019**, *31*, 1.
[31] W. Legrand, J.-Y. Chauleau, D. Maccariello, N. Reyren, S. Collin, K. Bouzehouane, N. Jaouen, V. Cros, A. Fert, *Sci. Adv* **2018**, *4*, 415.
[32] A. K. Nandy, N. S. Kiselev, S. Blügel, *Phys. Rev. Lett.* **2016**, *116*, 1.
[33] X. Zhang, Y. Zhou, M. Ezawa, *Nat. Commun.* **2016**, *7*, 1.
[34] Y. Liu, Y. Luo, Z. Qian, J. Zhu, *Chinese Phys. B* **2018**, *27*, 127503.
[35] J. L. Grab, A. E. Rugar, D. C. Ralph, *Phys. Rev. B* **2018**, *184424*, 1.
[36] L. A. C. M. E.Stebliy, A. G. Kolesnikov, A. V. Davydenko, A. V. Ognev, A. S. Samardak, *J. Appl. Phys.* **2015**, *117*, 3.
[37] L. Wang, Q. Feng, Y. Kim, R. Kim, K. H. Lee, S. D. Pollard, Y. J. Shin, H. Zhou, W. Peng, D. Lee, W. Meng, H. Yang, J. H. Han, M. Kim, Q. Lu, T. W. Noh, *Nat. Mater.* **2018**, *17*, 1087.
[38] H. Zhou, L. Wang, Y. Hou, Z. Huang, Q. Lu, W. Wu, *Nat. Commun.* **2015**, *6*, 1.
[39] Scheinfein, M. R. LLG micromagnetics simulator devoloped by M. R. Scheinfein, 31/12/2015, http://llgmicro.home.mindspring.com

# Figures

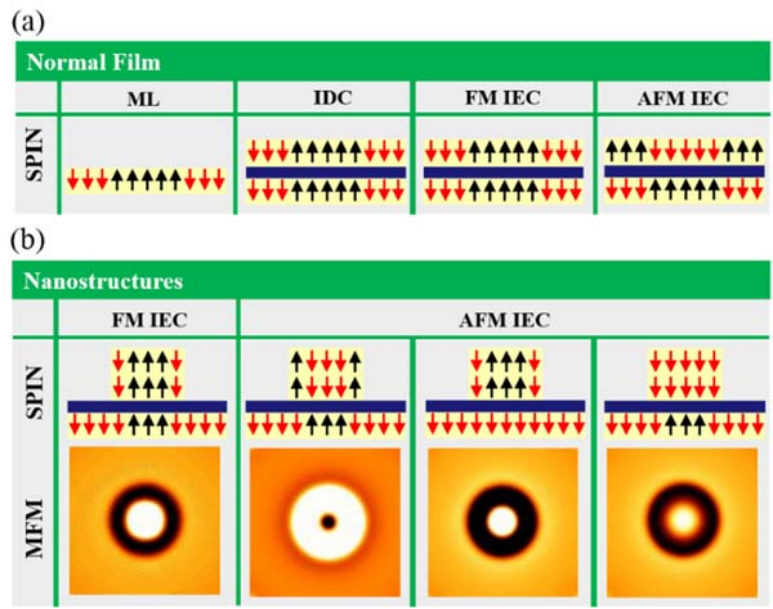

**Figure 1.** Cross-sections of skyrmion spin textures along the radial directions in (a) single- and multi-layer magnetic thin films and (b) nanostructured SAF multilayers (the upper panel). The black/red arrows represent the up/down out-of-plane magnetization. The corresponding simulated MFM images of skyrmion in nanostructured SAF multilayers are shown in the lower panel of (b).

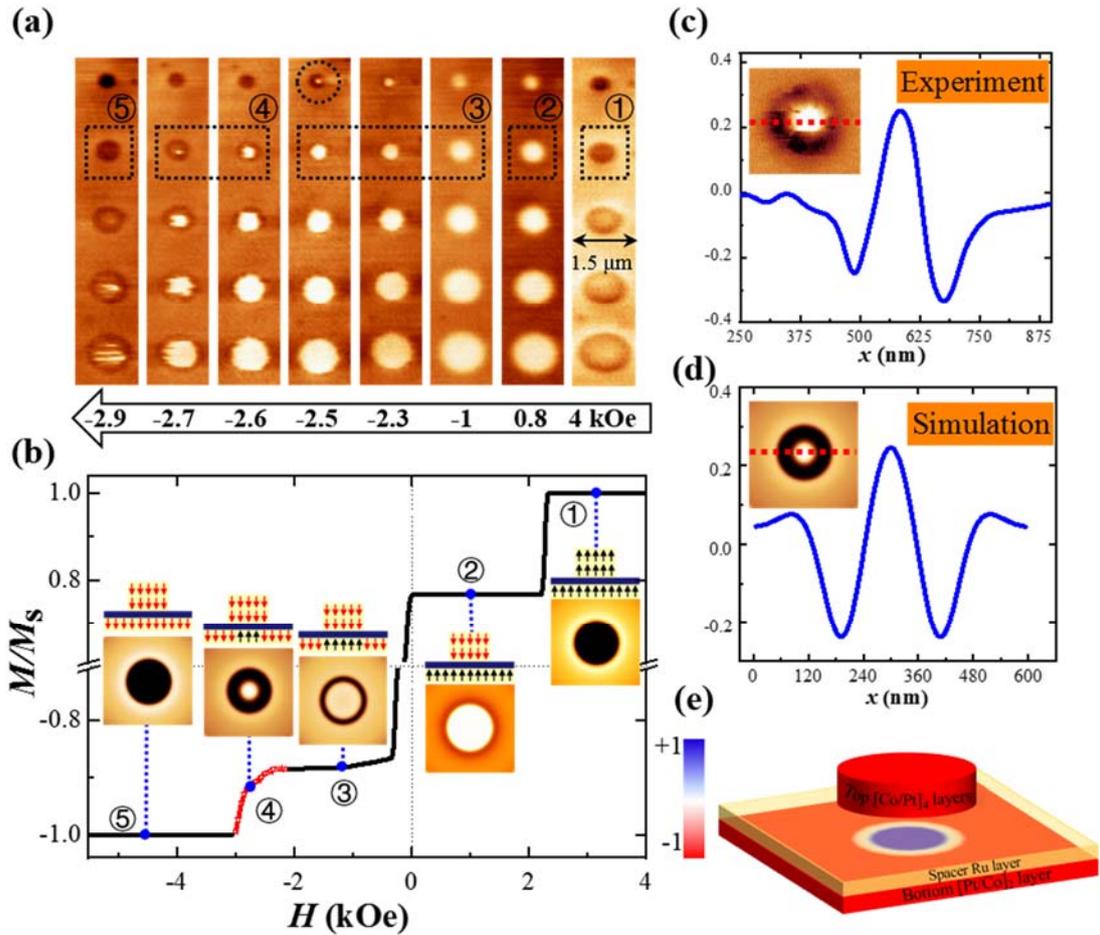

**Figure 2.** a) Measured room temperature MFM images of nanostructured SAF multilayer (diameters of top dots: 400 nm, 600 nm, 800 nm, 1000 nm, 1200 nm) at selected filed starting from positive saturation. b) Simulated *M-H* loop of nanostructured SAF multilayer with a top dot diameter of 400 nm. The insets show cross-sections of spin textures along the radial directions and the corresponding simulated MFM images at typical fields. Line cut along the dot radial direction of the (c) measured (as labelled by the dashed circle in (a)) and (d) simulated MFM images (as the inset at stage 4). e) Diagram of synthetic skyrmion configuration at stage 4 as labelled in (b).

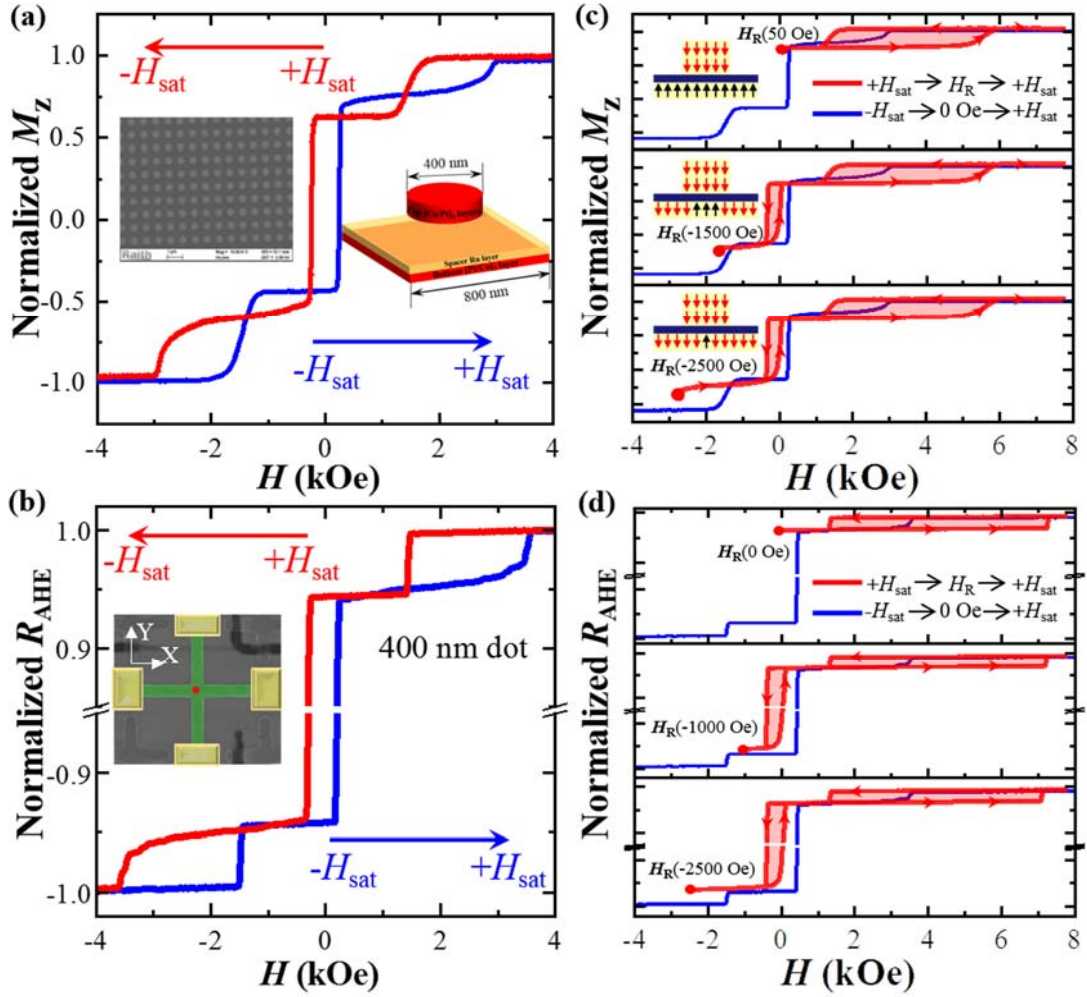

**Figure 3.** a) Normalized out-of-plane *M-H* loop of nanostructured SAF multilayer with a top dot diameter of 400 nm measured by Kerr microscopy at room temperature. Insert (left) is the SEM image of nanostructured SAF multilayer with a two-dimensional arrangement of the top 400 nm dot in a square lattice. Insert (right) is the schematic of one unit cell. b) Normalized anomalous Hall resistance loops of cross-structured SAF multilayer with a top dot diameter of 400 nm by magnetoresistance measurement. Inset is the SEM image of the Hall bar. The minor loops of normalized out-of-plane magnetization and anomalous Hall resistance are shown in (c) and (d). Blue line indicates the major loops and insets are the corresponding spin textures at the selected reversal fields $H_R$.

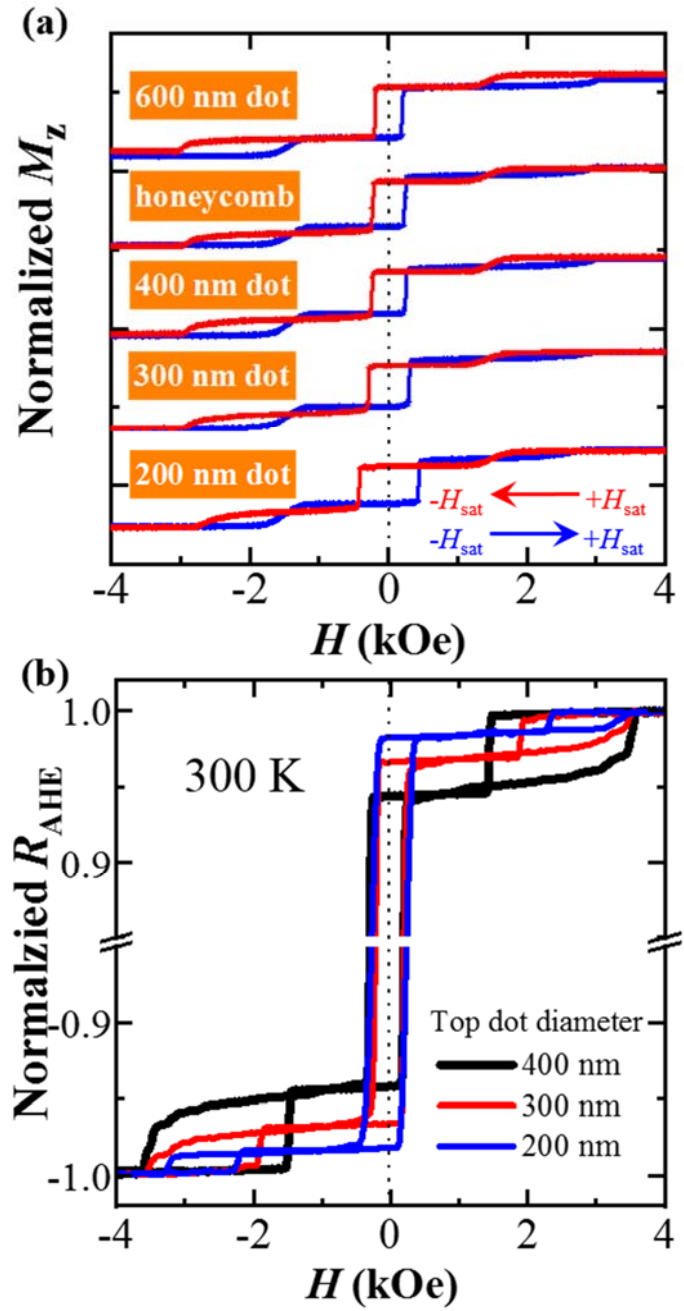

**Figure 4.** a) Normalized out-of-plane *M-H* loops of nanostructured SAF films with top dots of various diameters measured by MOKE at RT. b) Normalized $R_{AHE}$ loops of structured SAF films with top dots of various diameters measured by magnetoresistance transport.

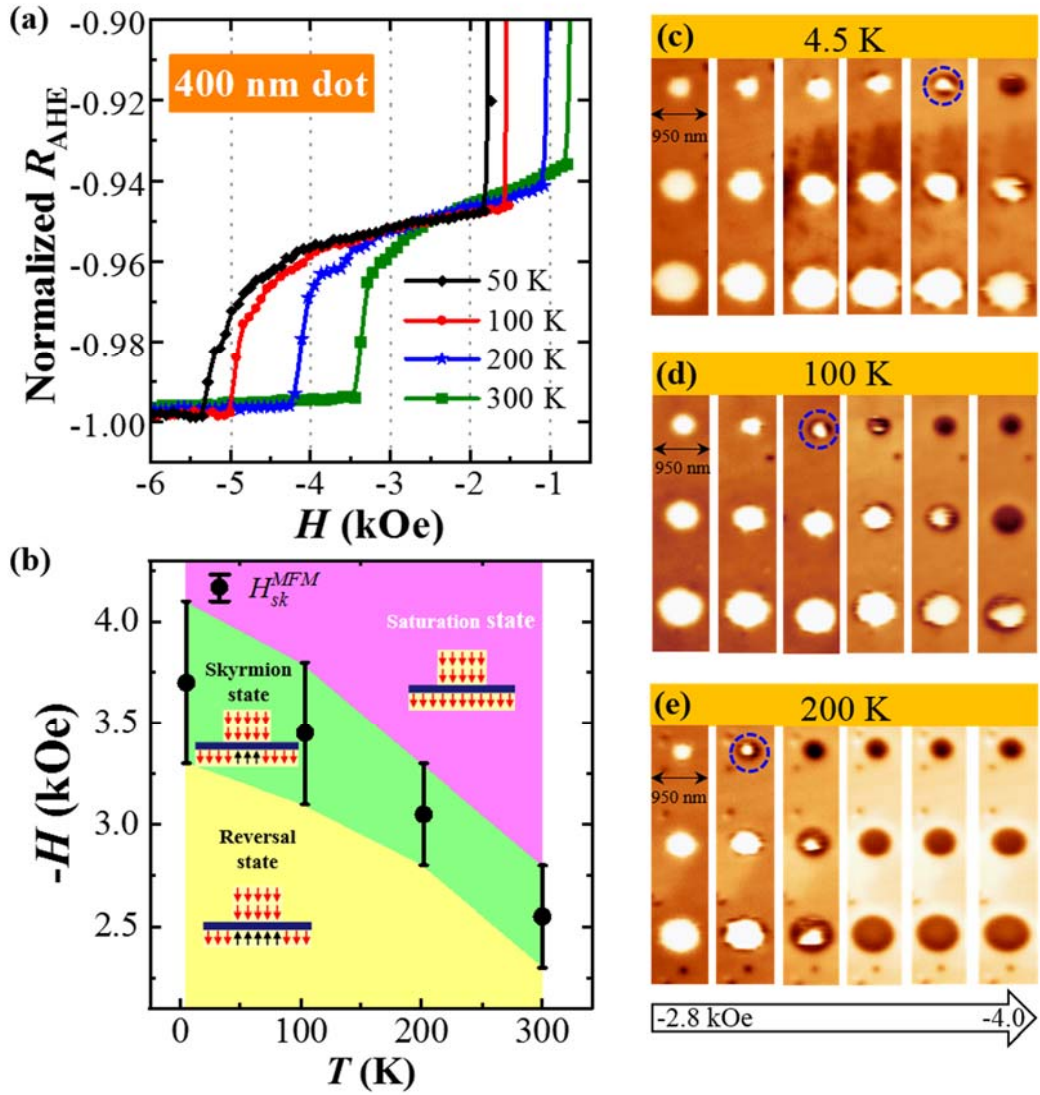

**Figure 5.** a) Fraction of the measured anomalous hall loops of nanostructured SAF film with a top dot of diameter 400nm at various temperatures. b) Field-temperature phase diagrams nanostructured SAF film with a top dot of diameter 400 nm derived from the MFM measurements. The phase diagram represents three different phases: Reversal phase (skyrmion size is equal to the physical size of the top dot), Skyrmion phase (skyrmion size is smaller than the physical size of the top dot), and Saturation phase (saturated along external field). Insets are the cross-sections of spin textures along the top dot radial directions. Low-temperature MFM images measured at c) 4.5 K, d) 100 K, and e) 200 K for the top dots with diameters of 400 nm, 600 nm, and 800 nm, respectively.

16